\documentclass[conference,a4paper]{IEEEtran}
\usepackage{latexsym}
\usepackage{graphicx}
\usepackage{amsfonts,amssymb,amsmath}
\usepackage{ctable} 
\usepackage{mathtools, cuted}
\usepackage{hyperref}
\usepackage[ruled,linesnumbered,noend]{algorithm2e}
\usepackage[T1]{fontenc}
\usepackage{cite}
\usepackage{subcaption}
\usepackage{comment}
\usepackage{diagbox}

\usepackage[a4paper, left=1.34 cm, right=1.34 cm, top=0.7in, bottom=4.15 cm]{geometry}

\usepackage{amsthm}
\usepackage{overpic}
\usepackage{steinmetz}
\usepackage{array}
\usepackage{url}
\usepackage{color}

\usepackage{algorithmicx}
\usepackage{algcompatible}

\usepackage{xcolor}
\usepackage{soul}

\theoremstyle{plain}

\newcommand{\vect}[1]{\mathbf{#1}}
\newcommand{\maximize}[1]{{\underset{{#1}}{\mathrm{maximize}}}}
\newcommand{\minimize}[1]{{\underset{{#1}}{\mathrm{minimize}}}}

\newcommand{\bl}[1]{\boldsymbol{#1}}
\newcommand{\mtr}[1]{\mathrm{#1}}
\newcommand*{\LongState}[1]{\STATE
\parbox[t]{0.9\linewidth-\algorithmicindent-\algorithmicindent}{#1\strut}}

\def\CN{\mathcal{N}_{\mathbb{C}}} 
\def\imagunit{\mathsf{j}} 

\def\T{\mathrm{T}}
\def\H{\mathrm{H}}

\IEEEoverridecommandlockouts

\title{A Novel Hybrid Precoder With Low-Resolution Phase Shifters and Fronthaul Capacity Limitation}
\author{Parisa Ramezani, Alva Kosasih, and Emil Bj\"{o}rnson\\
\IEEEauthorblockA{\textit{ Department of Computer Science, KTH Royal Institute of Technology, Stockholm, Sweden} \\  Email: \{parram, kosasih, emilbjo\}@kth.se}%
\thanks{This work was supported by the FFL18-0277 grant and SUCCESS project (FUS21-0026), funded by the Swedish Foundation for Strategic Research.}
\vspace{-3mm}}

\begin{document}
\maketitle

\begin{abstract}
In massive MIMO systems, fully digital precoding offers high performance but has significant implementation complexity and energy consumption, particularly at millimeter frequencies and beyond. Hybrid analog-digital architectures provide a practical alternative by reducing the number of radio frequency (RF) chains while retaining performance in spatially sparse multipath scenarios. However, most hybrid precoder designs assume ideal, infinite-resolution analog phase shifters, which are impractical in real-world scenarios. Another practical constraint is the limited fronthaul capacity between the baseband processor and array, implying that each entry of the digital precoder must be picked from a finite set of quantization labels. To minimize the sum rate degradation caused by quantized analog and digital precoders, we propose novel designs inspired by the sphere decoding (SD) algorithm. We demonstrate numerically that our proposed designs outperform traditional methods, ensuring minimal sum rate loss in hybrid precoding systems with low-resolution phase shifters and limited fronthaul capacity.
\end{abstract}
\vspace{-2mm}
\section{Introduction}
In massive multiple-input multiple-output (MIMO) systems operating in sub-6 GHz bands, fully digital precoding is commonly used. This means that each antenna is connected to a dedicated radio frequency (RF) chain. The significant implementation complexity and energy consumption of this approach at high frequencies (millimeter wave and beyond) with large bandwidths have led to the development of hybrid analog-digital architectures. These architectures utilize many analog phase shifters alongside a limited number of RF chains, offering an effective balance between performance and complexity.
Various hybrid precoding designs have been reported in the literature, where optimizing both the analog and digital precoders has resulted in performance close to that achievable by fully digital precoding in spatially sparse channels \cite{Sohrabi2015Hybrid,Liu2020Low,Zhan2021Interference,Wang2022A}.

When designing hybrid precoding, it is essential to consider the practical limitations of both analog and digital precoders.
Most research on hybrid precoder design assumes infinite-resolution phase shifters for analog precoders. However, practical phase shifters typically have a limited phase resolution. Several previous studies have proposed analog precoder designs that account for the low resolution of phase shifters \cite{Sohrabi2016Hybrid,ni2017near,Chen2018Low}. In these works, the resolution limitation of the analog phase shifters is initially ignored, and the optimal analog precoder is designed assuming infinite-resolution phase shifters. Subsequently, each optimized phase shift is mapped to the nearest value within the available set of low-resolution phase shifts. This approach results in a sub-optimal solution to the analog precoder design problem and leads to remarkable performance degradation, especially in the presence of interference and when the phase shifters have very low resolution. Further work on analog low-resolution precoding is needed. 

Another practical constraint is the fronthaul capacity limitation. A typical 5G base station (BS) consists of an advanced antenna system (AAS)  and a baseband unit (BBU), where a digital fronthaul connects AAS and BBU \cite{asplund2020advanced}. As the number of antennas and bandwidth increase, the amount of data to be exchanged between the BBU and the AAS grows enormously, leading to a fronthaul bottleneck. This limits the digital precoder design as the precoder entries must be compressed to fit within a limited number of bits. While fronthaul limitation has been investigated for fully digital precoders in \cite{khorsandmanesh2023optimized,ramezani2024joint}, no prior research on hybrid precoding has considered it.

Motivated by the practical limitations of analog and digital precoders and the lack of relevant research addressing these limitations, this paper proposes new designs that minimize the performance degradation caused by the limited bit resolution of the phase shifters and the fronthaul link. We employ a matrix decomposition approach to minimize the Euclidean distance between the optimized fully digital precoder and the hybrid precoder. To obtain the optimal analog and digital precoders under the aforementioned practical limitations, we introduce novel designs leveraging one of the well-known MIMO detection algorithms, namely, sphere decoding (SD) \cite{Agrell2002}. To the best of our knowledge, this is the first work to derive the optimal low-resolution analog and digital precoders in a hybrid precoding setup. We corroborate the effectiveness of the proposed designs through sum rate evaluations.  
\vspace{-1mm}
\section{System Model}
\vspace{-1mm}
We consider a multi-user MIMO system in which a BS, equipped with $N_{\mtr{T}}$ antennas and $M_{\mtr{T}}$ RF chains, serves $K$ single-antenna users over $S$ subcarriers. The number of RF chains is selected such that $K \leq M_{\mtr{T}} < N_{\mtr{T}}$ to ensure efficient operation of the hybrid precoding architecture. 
The downlink transmit symbol is first processed by the baseband digital precoder, $\vect{F}_{\mtr{BB}} \in \mathbb{C}^{M_{\mtr{T}} \times KS}$, and then the BS applies the fully-connected RF analog precoder, $\vect{F}_{\mtr{RF}} \in \mathbb{C}^{N_{\mtr{T}} \times M_{\mtr{T}}}$. Specifically, $\vect{F}_{\mtr{BB}} = [\vect{F}_{\mtr{BB},1},\ldots,\vect{F}_{\mtr{BB},K}]$, where $\vect{F}_{\mtr{BB},k} \in \mathbb{C}^{M_{\mtr{T}} \times S}$ is the BS digital precoder for the signal of user\,$k$ over all subcarriers. Therefore, the transmitted signal from the BS is $\Tilde{\vect{x}} = \vect{F}_{\mtr{RF}} \vect{F}_{\mtr{BB}} \vect{x}$, where $\vect{x} = [\vect{x}_1^\T,\ldots,\vect{x}_K^\T]^\T$, with $\vect{x}_k \in \mathbb{C}^S$ representing the signal vector intended for user\,$k$.
\vspace{-1.2mm}
\subsection{Discrete Set of Analog and Digital Precoding Entries}
We assume that the RF analog precoder is implemented using low-resolution phase shifters. Each entry of $\vect{F}_{\mtr{RF}}$ has a unit amplitude and belongs to a uniformly sampled discrete set determined by the resolution of the phase shifters. Hence, 
\begin{equation}
\label{eq:analog_precoder_set}
  \vect{F}_{\mtr{RF}}(n,m) \in \mathcal{D} = \left\{e^{\imagunit\frac{\ell \pi}{2^{b-1}}}: \ell =0,1,\ldots,2^b - 1 \right\},   
\end{equation}
where $\vect{F}_{\mtr{RF}}(n,m)$ is the entry in $n$th row and $m$th column of $\vect{F}_{\mtr{RF}}$, and $b$ is the resolution of the analog phase shifters. 

Furthermore, as the number of antennas and bandwidth increase, the data exchanged between the BBU and AAS grows significantly, leading to a fronthaul bottleneck. 
To mitigate this issue, a common approach is to transmit quantized data over the fronthaul, meaning that the entries of the digital precoder, $\vect{F}_{\mtr{BB}}$, must be selected from a discrete set rather than taking arbitrary values, to comply with the limited fronthaul capacity.
We define the set of digital precoding labels as \cite{jacobsson2017quantized,khorsandmanesh2023optimized,ramezani2024joint}
\begin{equation}
    \mathcal{B} = \{p_R + \imagunit p_I : p_R,p_I \in \mathcal{P}\}, 
\end{equation}
where $\mathcal{P}$ contains real-valued quantization labels, given by 
   $\mathcal{P} = \{p_0,p_1,\ldots,p_{L-1}\}$,
with $q = \log_2(L)$ representing the bit resolution of the fronthaul link. The entries of $\mathcal{P}$ are 
\begin{equation}
    p_i = \Delta \left(i - \frac{L-1}{2}\right),~~i = 0,\ldots,L-1, 
\end{equation}
where $\Delta$ depends on the statistical distribution of the precoding entries and must be selected to minimize the distortion between quantized and unquantized entries \cite{Hui2001,jacobsson2017quantized}. 
\vspace{-1.2mm}
\subsection{Sum Rate Maximization Problem}
Let $\vect{H} = [\vect{H}_1,\ldots,\vect{H}_K] \in \mathbb{C}^{N_{\mtr{T}} \times KS}$ denote the channel between the BS and the users, where $\vect{H}_k = \big[\vect{h}_k[1],\vect{h}_k [2],\ldots, \vect{h}_k [S]\big] \in \mathbb{C}^{N_{\mtr{T}} \times S}$ is the channel between the BS and user\,$k$ over all subcarriers, and $\vect{h}_k[s]$ is the channel between the BS and user\,$k$ on the $s$th subcarrier. 
The received signal at user\,$k$ on the $s$th subcarrier is 
\begin{align}
  y_k[s] &=  \vect{h}^\T_k[s] \vect{F}_{\mtr{RF}} \vect{F}_{\mtr{BB},k}[:,s] \vect{x}_k[s] \nonumber \\ &+ \sum_{i=1,i\neq k}^K \vect{h}_k^\T[s] \vect{F}_{\mtr{RF}} \vect{F}_{\mtr{BB},i}[:,s] \vect{x}_i[s] + n_k[s],
\end{align}
where $\vect{F}_{\mtr{BB},k}[:,s]$ denotes the $s$th 
 column of $\vect{F}_{\mtr{BB},k}$ and $n_k[s] \sim \CN (0,N_0)$ represents the independent additive Gaussian noise with power $N_0$.
The signal-to-interference-plus-noise ratio (SINR) at user\,$k$ on the $s$th subcarrier is
\begin{equation}
  \gamma_k[s]  = \frac{|\vect{h}^\T _k[s] \vect{F}_{\mtr{RF}}\vect{F}_{\mtr{BB},k}[:,s]|^2}{\sum_{i=1,i\neq k}|\vect{h}^\T 
 _k[s]\vect{F}_{\mtr{RF}} \vect{F}_{\mtr{BB},i}[:,s]|^2 + N_0}, 
\end{equation}
where it is assumed that the transmitted symbols and noise are mutually independent, and have unit power.
The achievable rate for user\,$k$ on the $s$th subcarrier is then calculated as $R_k[s] = \log_2(1+\gamma_k[s])$ when the user knows the channel.

We aim to design the discrete analog and digital precoders to maximize the sum rate. 
Consider the following problem:
\vspace{-2mm}
\begin{subequations}
\begin{align}
\maximize{\substack{\vect{F}_{\mtr{RF}} \in \mathcal{D}^{N_{\mtr{T}} \times 
 M_\mtr{T}},\\ \vect{F}_{\mtr{BB}} \in \mathcal{B}^{M_{\mtr{T}} \times KS}}}~~&\sum_{k=1}^K \sum_{s = 1}^S R_k[s] \\
   \mathrm{subject~to} \quad ~ &\label{eq:power_const}\sum_{k = 1}^{K}\big\|\vect{F}_{\mtr{RF}} \vect{F}_{\mtr{BB},k}[:,s]\big\|^2 \leq P,~~\forall s,
\end{align}
\end{subequations}
where \eqref{eq:power_const} represents the total transmit power constraint on each subcarrier and $\| \cdot \|$ denotes the Euclidean norm. This problem is challenging to solve because the objective function is not concave with respect to the optimization variables, the analog and digital precoders are coupled in both the objective function and constraint \eqref{eq:power_const}, and we have discrete constraints on analog and digital precoders. In the following sections, we apply a matrix decomposition approach to  minimize the Euclidean distance between the
 fully digital and hybrid precoders and employ an alternating optimization algorithm to decouple the analog and digital precoder optimization problems. 

\vspace{-0.5mm}
\section{Hybrid Precoder Design with Low-Resolution Phase Shifters}

To find the rate-maximizing hybrid precoder, we first consider a fully digital BS where each of $N_{\mtr{T}}$ antennas is connected to a separate RF chain. After finding the optimal precoder in this fully digital setup, we apply a matrix decomposition approach similar to \cite{ni2017near} to find a high-dimensional analog precoder and low-dimensional digital precoder such that their product approximates the optimized fully digital precoder as closely as possible. Here, we neglect the fronthaul capacity limitation and assume the digital precoder entries can take arbitrary values. In the next section, we will design the digital precoder under the fronthaul capacity constraint. 

Assuming that $\vect{F}_{\mtr{FD}}^\star \in \mathbb{C}^{N_{\mtr{T}} \times KS}$ is the optimal fully digital precoder, we formulate the following problem to optimize the analog precoder $\vect{F}_{\mtr{RF}}$ and digital precoder $\vect{F}_{\mtr{BB}}$:
\begin{subequations}
\label{eq:difference_minimization}
\begin{align}
    \minimize{\substack{\vect{F}_{\mtr{RF}} \in \mathcal{D}^{N_{\mtr{T}} \times 
 M_\mtr{T}},\\ \vect{F}_{\mtr{BB}} \in \mathbb{C}^{M_{\mtr{T}} \times KS}}} ~&\|\vect{F}_{\mtr{FD}}^\star - \vect{F}_{\mtr{RF}}\vect{F}_{\mtr{BB}}\|_F^2 \\
    \mathrm{subject~to} \,\,\, ~& \|\vect{F}_{\mtr{RF}}\vect{F}_{\mtr{BB}}\|_F^2 = \|\vect{F}_{\mtr{FD}}^\star\|_F^2,
\end{align}
\end{subequations}
where $\|\cdot\|_F$ denotes the Frobenius norm. 
With the fully digital precoder $\vect{F}_{\mtr{FD}} = [\vect{F}_{\mtr{FD},1},\ldots, \vect{F}_{\mtr{FD},K}]$ at the BS, the SINR at user\,$k$ on the $s$th subcarrier is given by 
\begin{equation}
  \Tilde{\gamma}_k[s]  = \frac{|\vect{h}^\T_k[s] \vect{F}_{\mtr{FD},k}[:,s]|^2}{\sum_{i=1,i\neq k}|\vect{h} ^\T  _k[s]\vect{F}_{\mtr{FD},i}[:,s]|^2 + N_0}, 
\end{equation}
where $\vect{F}_{\mtr{FD},k}[:,s]$ is the precoder for user\,$k$'s signal on the $s$th subcarrier. The achievable rate of user\,$k$ is $\sum_{s = 1}^S\Tilde{R}_k[s] = \sum_{s=1}^S\log_2(1+\Tilde{\gamma}_k[s])$ in this case, and the sum rate maximization problem is formulated as 
\begin{subequations}
\label{eq:fully_digital_problem}
\begin{align}
\maximize{\vect{F}_{\mtr{FD}} \in \mathbb{C}^{N_{\mtr{T}} \times KS}  }\, &\sum_{k=1}^K \sum_{s = 1}^S \Tilde{R}_k[s],\\ \mtr{subject~to}\,\, &\sum_{k=1}^K \|\vect{F}_{\mtr{FD},k}[:,s]\|^2 \leq P,~~\forall s.
\end{align}
\end{subequations}
Problem \eqref{eq:fully_digital_problem} can be solved using the classical WMMSE approach. In this approach, the sum rate maximization problem is formulated as a weighted sum mean squared error minimization problem, and an iterative algorithm is employed to guarantee convergence to at least a local optimum of the original sum rate maximization problem. The details of the WMMSE approach can be found in \cite{Shi2011}. 

We now return to problem \eqref{eq:difference_minimization}, aiming to find the analog and digital precoders such that their product closely approximates the  fully digital precoder $\vect{F}_{\mtr{FD}}^\star$. We propose an iterative approach where the precoders are alternately optimized until a satisfactory convergence is achieved. We first fix the analog precoder and solve problem \eqref{eq:difference_minimization} for the digital precoder. Then, with the digital precoder fixed, we propose a novel method, inspired by a classical MIMO detection algorithm, to optimize the low-resolution analog precoder. 
\vspace{-1mm}
\subsection{Optimizing the Digital Precoder}
 We fix the analog precoder $\vect{F}_{\mtr{RF}}$ and formulate the digital precoder optimization problem as
\begin{subequations}
\label{eq:dig_precoder_opt}
\begin{align}
    \minimize{\vect{F}_{\mtr{BB}} \in \mathbb{C} ^{M_{\mtr{T}} \times KS}} ~&\|\vect{F}_{\mtr{FD}}^\star - \vect{F}_{\mtr{RF}}\vect{F}_{\mtr{BB}}\|_F^2 \\
    \mathrm{subject~to}~& \|\vect{F}_{\mtr{RF}}\vect{F}_{\mtr{BB}}\|_F^2 = \|\vect{F}_{\mtr{FD}}^\star\|_F^2. \label{eq:power_equal}
\end{align}
\end{subequations}
We initially disregard the constraint  \eqref{eq:power_equal}. This constraint will be enforced at the end of the iterative process to ensure that the total power constraint is satisfied. In the absence of \eqref{eq:power_equal}, the solution to \eqref{eq:dig_precoder_opt} is readily obtained as 
\begin{equation}
\label{eq:digital_precoder}
    \check{\vect{F}}_{\mtr{BB}} =  \left(\vect{F}_{\mtr{RF}} \right)^\dagger \vect{F}_{\mtr{FD}}^\star,
\end{equation}where $(\cdot)^\dagger$ indicates the pseudoinverse. 
\vspace{-2mm}
\subsection{Optimizing the Analog Precoder }
Next, for a fixed digital precoder $\vect{F}_{\mtr{BB}}$, the problem of optimizing the analog precoder is formulated as
\begin{align}
 \label{eq:analog_prec_optimization}\minimize{\vect{F}_{\mtr{RF}} \in \mathcal{D}^{N_{\mtr{T}} \times M_{\mtr{T}}}}~& \|\vect{F}_{\mtr{FD}}^\star - \vect{F}_{\mtr{RF}}\vect{F}_{\mtr{BB}}\|_F^2. 
\end{align}
Since the Frobenius norm of a matrix equals that of its transpose, the objective function in \eqref{eq:analog_prec_optimization} can be re-written as 
\begin{equation}
  \|\vect{F}_{\mtr{FD}}^\star - \vect{F}_{\mtr{RF}}\vect{F}_{\mtr{BB}}\|_F^2 = \|\vect{F}_{\mtr{FD}}^{\star \T} - \vect{F}_{\mtr{BB}}^\T \vect{F}_{\mtr{RF}}^\T\|_F^2. 
\end{equation}
For notational simplicity, we define $\vect{A} \triangleq \vect{F}_{\mtr{FD}}^{\star \T} \in \mathbb{C}^{KS\times N_{\mtr{T}}}$, $\vect{B} \triangleq \vect{F}_{\mtr{BB}}^\T \in \mathbb{C}^{KS\times M_{\mtr{T}}}$, and $\vect{X} \triangleq \vect{F}_{\mtr{RF}}^\T \in \mathbb{C}^{M_{\mtr{T}}\times N_{\mtr{T}}}$. The analog precoder optimization problem is re-formulated as 
\begin{align}
\label{eq:analog_prec_optimization2}
 \minimize{\vect{X} \in \mathcal{D}^{M_{\mtr{T}} \times N_{\mtr{T}}}}~& \|\vect{A} - \vect{BX}\|_F^2.
\end{align}
Since the square of the Frobenius norm of a matrix is the sum of the squares of the Euclidean norms of its columns, the problem in \eqref{eq:analog_prec_optimization2} is further re-formulated as 
\begin{align}
\minimize{{\vect{X}} \in \mathcal{D}^{M_{\mtr{T}} \times N_{\mtr{T}}}}~&\sum_{n=1}^{N_\mtr{T}}\|\vect{a}_n - \vect{B}\vect{x}_n\|^2,
\end{align}
where $\vect{a}_n$ and $\vect{x}_n$ are the $n$th columns of $\vect{A}$ and $\vect{X}$, respectively. This problem can be divided into $N_{\mtr{T}}$ separate sub-problems:
\begin{align}
 \label{eq:separate_subproblems}\minimize{\vect{x}_n \in \mathcal{D}^{M_{\mtr{T}}}}\,\|\vect{a}_n - \vect{B}\vect{x}_n\|^2.  
\end{align}
We notice that \eqref{eq:separate_subproblems} has the same form as MIMO detection problems, where $\vect{a}_n$ and $\vect{B}$ resemble the received signal vector and the channel matrix, $\vect{x}_n$ is akin to the transmitted signal to be detected, and $\mathcal{D}$ represents the signal constellation. 

The optimal MIMO detection solution is obtained by the maximum likelihood approach which performs an exhaustive search over all possible vectors to find the global optimum. However, such a brute force algorithm incurs very high computational complexity. 
A more efficient alternative is SD, which significantly reduces the complexity without sacrificing optimality. SD narrows down the search space by only considering the candidate vectors that lie within a certain radius from the received signal vector, thereby reducing the number of required searches and computations. In this paper, we use the SD method to solve \eqref{eq:separate_subproblems} optimally. 
To apply the SD method,  we need a reformulation of the objective function as 
\begin{equation}
  \|\vect{a}_n - \vect{B}\vect{x}_n\|^2 = \|\vect{d}_n - \vect{R}\vect{x}_n\|^2 + \vect{a}_n^\H \vect{a}_n - \vect{d}_n^\H \vect{d}_n,  
\end{equation}
where $\vect{R}$ is an upper-triangular matrix obtained via the QR decomposition of $\vect{B}$ or the Cholesky decomposition of $\vect{B}^\H \vect{B}$ such that $\vect{R}^\H \vect{R} = \vect{B}^\H \vect{B}$, and $\vect{d}_n = (\vect{a}_n^\H \vect{B} \vect{R}^{-1})^\H$. 
As $\vect{R}$ is an upper-triangular matrix, we can optimize the analog precoder by solving the following problem using the SD method:
\begin{align}
\label{eq:SD_problem}
  \minimize{\vect{x}_n \in \mathcal{D}^{M_{\mtr{T}}}}\, \|\vect{d}_n - \vect{R}\vect{x}_n\|^2. 
\end{align}
Among various SD algorithms, we use the Schnorr-Euchner SD (SESD) \cite{Agrell2002} because its zig-zag search for the optimal solution reduces the number of explored vectors compared to other SD algorithms. Denoting the optimal solution of \eqref{eq:SD_problem} by $\check{\vect{x}}_n$, the optimal analog precoder given the digital precoder is 
\begin{equation}
\label{eq:analog_precoder}
   \check{\vect{F}}_{\mtr{RF}} = [\check{\vect{x}}_1,\check{\vect{x}}_2,\ldots,\check{\vect{x}}_{N_{\mtr{T}}}]^\T. 
\end{equation}

\subsection{Overall Algorithm}
The digital precoder and the analog precoder are iteratively optimized, as discussed in the previous subsections, until the relative change in the objective function falls below a predefined threshold. Once convergence is achieved, the digital precoder must be normalized to ensure that the constraint in \eqref{eq:power_equal} is satisfied. Denoting the optimized analog precoder after convergence by $\vect{F}_{\mtr{RF}}^\star$, the final digital precoder is obtained as 
\begin{equation}
    \vect{F}_{\mtr{BB}}^\star = \frac{\|\vect{F}_{\mtr{FD}}^\star\|_F}{\|\vect{F}_{\mtr{RF}}^\star  \check{\vect{F}}_{\mtr{BB}}\|_F} \check{\vect{F}}_{\mtr{BB}}.
\end{equation}

The proposed hybrid precoder design is summarized in Algorithm~\ref{Alg:Hybrid_Precoding}. There are several methods for initializing the analog precoder. One approach is random initialization, where each entry of the analog precoder has a unit modulus, with phases independently selected from a uniform distribution over 
 $[0,2\pi)$. Another way is to initialize it as 
\begin{equation}
  \vect{F}_{\mtr{RF}} = \exp\left(\imagunit\arg(\Tilde{\vect{U}}_{\mtr{FD}} \Tilde{\bl{\Sigma}}_{\mtr{FD}}) \right),  
\end{equation}
where $\Tilde{\bl{\Sigma}}_{\mtr{FD}} \in \mathbb{C}^{M_{\mtr{T}} \times M_{\mtr{T}}}$ is a diagonal matrix having the $M_{\mtr{T}}$ largest singular values of $\vect{F}^\star_{\mtr{FD}}$ on its diagonal and $\tilde{\vect{U}}_{\mtr{FD}} \in \mathbb{C}^{N_{\mtr{T}} \times M_{\mtr{T}}}$ consists of the corresponding left singular vectors. Our simulations have shown that the second approach generally yields higher sum rates; thus, we have used the second initialization method in Section~\ref{sec:NumRes}.

\begin{algorithm}
\caption{Hybrid Precoder Design with Low-Resolution Analog Precoder}
\label{Alg:Hybrid_Precoding}
\begin{algorithmic}[1]
\STATEx {\textbf{Inputs}: Channel $\vect{H}$, an initial analog precoder $\vect{F}_{\mtr{RF}}$}
\STATE{Find the optimal fully digital precoder by solving problem \eqref{eq:fully_digital_problem} using the WMMSE algorithm}
\REPEAT
\LongState{Compute the optimal digital precoder $\check{\vect{F}}_{\mtr{BB}}$ in \eqref{eq:digital_precoder} for a given analog precoder  }
\LongState{Compute the optimal analog precoder $\check{\vect{F}}_{\mtr{RF}}$ in \eqref{eq:analog_precoder}, for a given the digital precoder, by solving problem \eqref{eq:SD_problem} using SESD for all $n = 1,\ldots,N_{\mtr{T}}$}
\UNTIL{The relative change of the objective function is \,less than a predefined threshold}
\STATE{Set $\vect{F}_{\mtr{RF}}^\star = \check{\vect{F}}_{\mtr{RF}},\,\vect{F}_{\mtr{BB}}^\star = \frac{\|\vect{F}_{\mtr{FD}}^\star\|_F}{\|\vect{F}_{\mtr{RF}}^\star  \check{\vect{F}}_{\mtr{BB}}\|_F} \check{\vect{F}}_{\mtr{BB}}$}
 \end{algorithmic}
\end{algorithm}
\vspace{-3mm}
\section{Optimizing the Digital Precoder Under the Fronthaul Capacity Limitation}

In the presence of a fronthaul capacity limitation, we need to revisit problem \eqref{eq:dig_precoder_opt} and solve it by taking into account the finite-resolution constraint on the precoding entries. The new digital precoder optimization problem is formulated as 
\begin{subequations}
\label{eq:dig_prec_opt_fronthaul}
\begin{align}
    \minimize{\vect{F}_{\mtr{BB}} \in \mathcal{B}^{M_{\mtr{T}} \times KS}} ~&\|\vect{F}_{\mtr{FD}}^\star - \vect{F}_{\mtr{RF}}\vect{F}_{\mtr{BB}}\|_F^2 \\
    \mathrm{subject~to} \,\, ~& \sum_{k = 1}^{K}\|\vect{F}_{\mtr{RF}} \vect{F}_{\mtr{BB},k}[:,s]\|^2 \leq P,~~\forall s. \label{eq:pow_const_subcarrier}
\end{align}
\end{subequations}
Note that here, we cannot ensure $\|\vect{F}_{\mtr{RF}}\vect{F}_{\mtr{BB}}\|_F^2 = \|\vect{F}_{\mtr{FD}}^\star\|_F^2$ because of the discrete constraint on the digital precoder entries.  We can solve problem \eqref{eq:dig_prec_opt_fronthaul} using a Lagrange duality method, where the Lagrangian is given by
\begin{align}
 \mathcal{L}(\vect{F}_{\mtr{BB}},\bl{\mu}) &=  \|\vect{F}_{\mtr{FD}}^\star - \vect{F}_{\mtr{RF}}\vect{F}_{\mtr{BB}}\|_F^2 \nonumber \\&+ \sum_{s = 1}^S \mu_s \left(\sum_{k = 1}^{K}\|\vect{F}_{\mtr{RF}} \vect{F}_{\mtr{BB},k}[:,s]\|^2  - P\right). 
\end{align}
$\bl{\mu} = [\mu_1,\ldots,\mu_S]$ is the set of non-negative Lagrange multipliers associated with the power constraints \eqref{eq:pow_const_subcarrier}. We thus need to solve the following problem to obtain the dual function:
\begin{equation}
   \minimize{\vect{F}_{\mtr{BB}} \in \mathcal{B}^{M_{\mtr{T}} \times KS}}~ \mathcal{L}(\vect{F}_{\mtr{BB}},\bl{\mu}),
\end{equation}
which can be separated into $S$ sub-problems (one per subcarrier), with sub-problem $s$ being expressed as 
\begin{align}
\label{eq:S_subproblems}
\minimize{\{\vect{b}_{k,s}\}_{k=1}^K \in \mathcal{B}^{M_{\mtr{T}}}}~&\sum_{k = 1}^K \Big(\big\|\vect{a}_{k,s} - \vect{F}_{\mtr{RF}} \vect{b}_{k,s}\big\|^2 \nonumber \\& +\mu_s \big\|\vect{F}_{\mtr{RF}} \vect{b}_{k,s}\big\|^2\Big),     
\end{align}
where $\vect{a}_{k,s} \triangleq \vect{F}^\star_{\mtr{FD},k}[:,s] $ and $\vect{b}_{k,s} \triangleq \vect{F}_{\mtr{BB},k}[:,s]$. Problem \eqref{eq:S_subproblems} can be further divided into $K$ independent sub-problems, where sub-problem $k$ is
\begin{align}
\label{eq:K_subproblems}
  \minimize{\vect{b}_{k,s} \in \mathcal{B}^{M_{\mtr{T}}}}~&(\mu_s + 1)\vect{b}_{k,s}^\H \vect{F}^\H_{\mtr{RF}} \vect{F}_{\mtr{RF}} \vect{b}_{k,s} \nonumber \\ &- \vect{a}_{k,s}^\H \vect{F}_{\mtr{RF}}\vect{b}_{k,s}  - \vect{b}_{k,s}^\H \vect{F}^\H_{\mtr{RF}}\vect{a}_{k,s}.
\end{align}
We can re-write the objective function of \eqref{eq:K_subproblems} as 
\begin{equation}
  \|\Tilde{\vect{d}}_{k,s} - \Tilde{\vect{R}}_s \vect{b}_{k,s}\|^2 - \Tilde{\vect{d}}_{k,s}^\H  \Tilde{\vect{d}}_{k,s},
\end{equation}
with $\Tilde{\vect{R}}_s$ being an upper-triangular matrix obtained from the Cholesky decomposition of $(\mu_s + 1) \vect{F}^\H_{\mtr{RF}} \vect{F}_{\mtr{RF}}$, i.e., $\tilde{\vect{R}}_s^\H \tilde{\vect{R}}_s = (\mu_s + 1) \vect{F}^\H_{\mtr{RF}} \vect{F}_{\mtr{RF}}$, and $\tilde{\vect{d}}_{k,s} = (\vect{a}_{k,s}^\H \vect{F}_{\mtr{RF}}\tilde{\vect{R}}_s^{-1})^\H$. Accordingly, SESD can be utilized to find the digital precoder entries for user\,$k$ on the $s$th subcarrier by solving the following problem:
\begin{equation}
\label{eq:SD_dig_prec}
    \minimize{\vect{b}_{k,s} \in \mathcal{B}^{M_{\mtr{T}}}}~ \|\Tilde{\vect{d}}_{k,s} - \Tilde{\vect{R}}_s \vect{b}_{k,s}\|^2, 
\end{equation}
which is similar to the analog precoder optimization problem in \eqref{eq:SD_problem}. Since the real and imaginary parts of the digital precoder entries are independent and the same set of quantization labels $\mathcal{P}$ is utilized for both parts, we can express problem \eqref{eq:SD_dig_prec} in an equivalent real-valued form. To this end, we define
\begin{equation}
\begin{aligned}
&\tilde{\vect{d}}^r_{k,s} = \begin{bmatrix}
    \Re(\tilde{\vect{d}}_{k,s}) \\
     \Im(\tilde{\vect{d}}_{k,s}),
    \end{bmatrix},~ \tilde{\vect{b}}^r_{k,s} = \begin{bmatrix}
    \Re(\vect{b}_{k,s}) \\
     \Im(\vect{b}_{k,s}) 
    \end{bmatrix}, \\
   & \tilde{\vect{R}}^r_s =
  \begin{bmatrix}
    \Re(\tilde{\vect{R}}_s) & -\Im(\tilde{\vect{R}}_s)  \\
    \Im(\tilde{\vect{R}}_s)  & \Re(\tilde{\vect{R}}_s)
  \end{bmatrix},
\end{aligned}
\end{equation}
and re-formulate \eqref{eq:SD_dig_prec} as
\begin{equation}
    \label{eq:SD_dig_prec_real}
    \minimize{\tilde{\vect{b}}^r_{k,s} \in \mathcal{P}^{2M_{\mtr{T}}}}~ \|\Tilde{\vect{d}}^r_{k,s} - \Tilde{\vect{R}}^r_s \tilde{\vect{b}}^r_{k,s}\|^2. 
\end{equation}
Representing by $\check{\vect{b}}^r_{k,s}$ the solution to \eqref{eq:SD_dig_prec_real} obtained via SESD, the digital precoder under the fronthaul limitation for user\,$k$ on the $s$th subcarrier will be given by
\begin{equation}
\check{\vect{b}}_{k,s}(\mu_s) = \check{\vect{b}}^r_{k,s}[:,1:M_{\mtr{T}}] + \imagunit\check{\vect{b}}^r_{k,s}[:,M_{\mtr{T}}+1:2M_{\mtr{T}}].
\end{equation}
The optimal value of $\mu_s$, which satisfies the power constraint \eqref{eq:pow_const_subcarrier} near equality, can be found via bisection search.  Finally, the optimal digital precoder for a given analog precoder is
\begin{equation}
    \check{\vect{F}}_{\mtr{BB}}= \left[\check{\vect{F}}_{\mtr{BB},1},\ldots,\check{\vect{F}}_{\mtr{BB},K}\right],
\end{equation}
where $\check{\vect{F}}_{\mtr{BB},k}= \left[\check{\vect{b}}_{k,1}(\mu_1^\star),\ldots,\check{\vect{b}}_{k,S}(\mu_S^\star)\right]$
is the optimal digital precoder for user\,$k$, with $\mu_s^\star$ being the optimal Lagrange multiplier for the power constraint on the $s$th subcarrier. Algorithm~\ref{Alg:Hybrid_Precoding2} summarizes the hybrid precoder design in the presence of fronthaul capacity limitation. 

\begin{algorithm}
\caption{Hybrid Precoder Design with Low-Resolution Analog Precoder and Fronthaul Limitation}
\label{Alg:Hybrid_Precoding2}
\begin{algorithmic}[1]
\STATEx {\textbf{Inputs}: Channel $\vect{H}$, an initial analog precoder $\vect{F}_{\mtr{RF}}$, set of quantization labels $\mathcal{P}$}
\STATE{Find the optimal fully digital precoder by solving problem \eqref{eq:fully_digital_problem} using the WMMSE algorithm}
\REPEAT
\FOR{$s = 1:S$}
\REPEAT
\STATE{Solve \eqref{eq:SD_dig_prec_real} using SESD for all $k = 1,\ldots,K$ }
\STATE{Update $\mu_s$ using the bisection method}
\UNTIL{ $\left|\sum_{k = 1}^{K}\|\check{\vect{F}}_{\mtr{RF}} \check{\vect{F}}_{\mtr{BB},k}[:,s]\|^2  - P\right|$ is less than ~~a threshold or the bisection interval becomes very small}
\ENDFOR
\STATE{Set $\check{\vect{F}}_{\mtr{BB}}= \left[\check{\vect{F}}_{\mtr{BB},1},\ldots,\check{\vect{F}}_{\mtr{BB},K}\right]$}
\LongState{Compute the optimal analog precoder $\check{\vect{F}}_{\mtr{RF}}$ in \eqref{eq:analog_precoder}, for a given digital precoder, by solving problem \eqref{eq:SD_problem} using SESD for all $n = 1,\ldots,N_{\mtr{T}}$}
\UNTIL{The relative change of the objective function is \,less than a threshold}
\STATE{Set $\vect{F}_{\mtr{RF}}^\star = \check{\vect{F}}_{\mtr{RF}},\,\vect{F}_{\mtr{BB}}^\star =  \check{\vect{F}}_{\mtr{BB}}$} 
 \end{algorithmic}
\end{algorithm}
\vspace{-1mm}
\section{Numerical Results}
\label{sec:NumRes}
In this section, we evaluate the proposed hybrid precoder designs through Monte Carlo simulations. 
 First, we demonstrate convergence for the proposed algorithms and then compare the sum rate against benchmark schemes.

The following setup is used for all simulations: A BS with $N_{\mtr{T}} = 64$ half-wavelength-spaced antennas in the form of a uniform linear array (ULA), and equipped with $M_{\mtr{T}} = 8$ RF chains, serves $K = 2$ users on $S = 16$ subcarriers. 
The channel between the BS and user\,$k$ on the $s$th subcarrier is $\vect{h}_k[s] = \sum_{\ell = 0}^T \bar{\vect{h}}_k[\ell]e^{-\imagunit 2\pi \ell s/S}$, where $T$ is the number of channel taps and $\bar{\vect{h}}_k[\ell] \in \mathbb{C}^{N_{\mtr{T}}}$ is the time-domain channel at tap $\ell$ \cite{bjornson2024introduction}. We consider a Rician fading channel with $4$ taps, i.e., $T = 3$. The first tap is the line-of-sight (LoS) path and other taps are non-LoS (NLoS) i.i.d. Rayleigh fading channels. Specifically, 
\begin{align}
\label{eq:channel_models}
 &\bar{\vect{h}}_k[0] = \sqrt{\frac{\kappa}{\kappa + 1}} \sqrt{\beta}\,\bl{\mathfrak{a}}(\varphi_k), \nonumber \\ &\bar{\vect{h}}_k[\ell] = \sqrt{\frac{1}{\kappa+1}}\sqrt{\beta}\, \bar{\vect{h}}_{k,\mtr{i.i.d.}},~\ell \neq 0,  
\end{align}
where $\kappa$ is the Rician factor, set as $10\,$dB. $\beta$ is the pathloss given by $\beta = -22\log_{10}(d_k/1\,\mtr{m}) - 28 - 20\log_{10}(f_c/1\,\mtr{GHz})$ based on the 3GPP pathloss model \cite[Table B.1.2.1-1]{3gpp}, with $f_c$ being the carrier frequency and $d_k$ being the distance between the BS and user\,$k$. Additionally, $\varphi_k$ in \eqref{eq:channel_models} is the angle of departure from the BS towards user\,$k$ and $\bl{\mathfrak{a}}(\cdot)$ represents the BS array response vector. Finally, the entries of $\bar{\vect{h}}_{k,\mtr{i.i.d.}}$ are $\CN(0,1)$-distributed.
The users are uniformly distributed around the BS such that $\varphi_k \sim \mathcal{U}[-\frac{\pi}{3},\frac{\pi}{3})$ and  $d_k \sim \mathcal{U}[100\,\mtr{m},200\,\mtr{m}],\,\forall k$. The carrier frequency is $f_c = 28\,$GHz. The noise power spectral density is  $-164\,$dBm/Hz and the bandwidth per subcarrier is $10\,$MHz. Furthermore, the total powers reported in the figures represent the overall transmit power that is equally divided among the subcarriers. 

\begin{figure}
\centering
\subfloat[Convergence of Algorithm~\ref{Alg:Hybrid_Precoding}]
{\includegraphics[width=0.86\columnwidth]{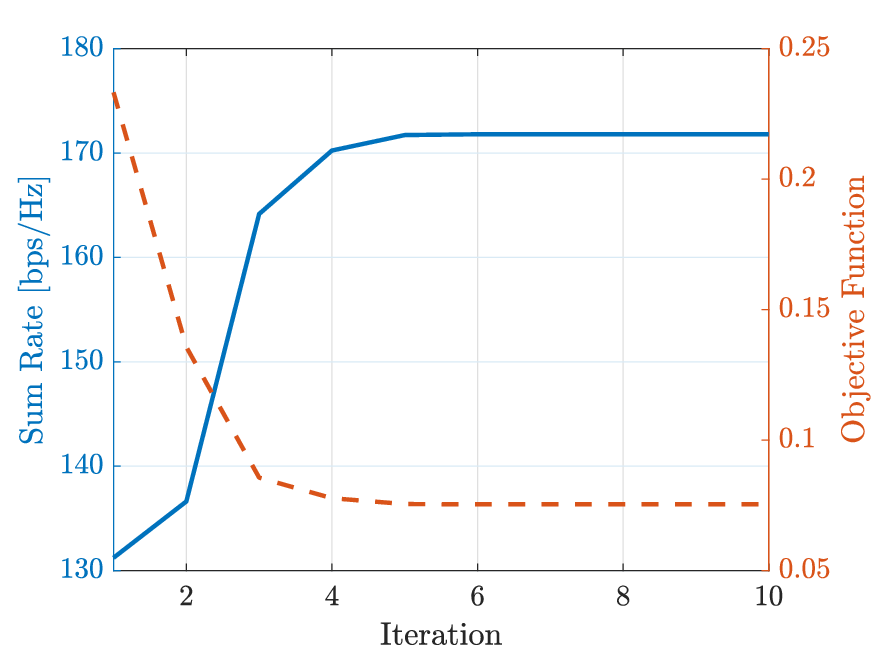} \label{fig:convergence1}} \hfill
\centering
\subfloat[Convergence of Algorithm~\ref{Alg:Hybrid_Precoding2}]
{\includegraphics[width=0.86\columnwidth]{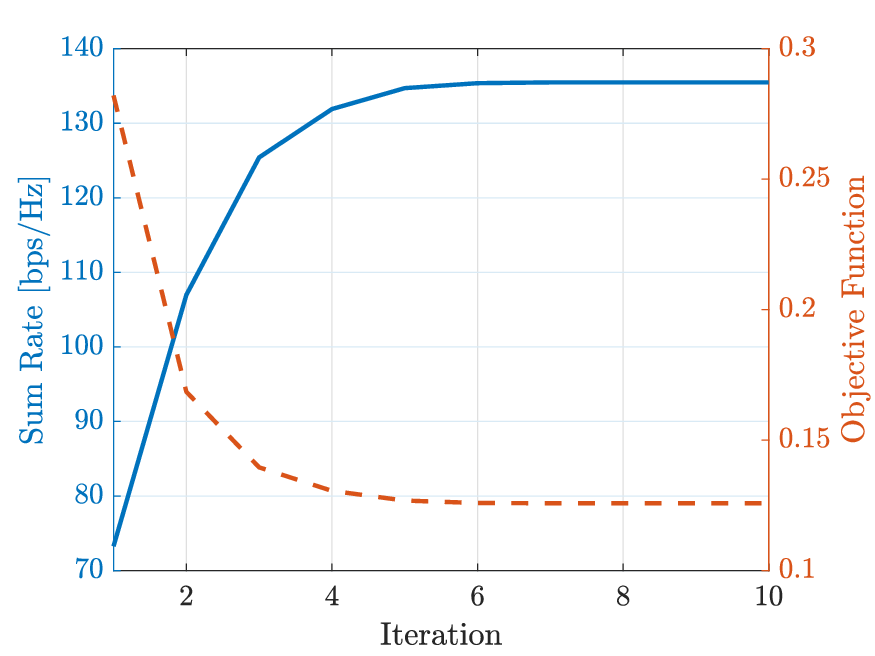} \label{fig:convergence2}} 
\caption{Convergence behavior of the proposed hybrid precoder designs, with and without fronthaul limitation.}
\label{fig:convergence}
\vspace{-4mm}
\end{figure}
Figs.~\ref{fig:convergence1} and \ref{fig:convergence2} show the convergence behavior of the proposed hybrid precoder designs in Algorithm~\ref{Alg:Hybrid_Precoding} and Algorithm~\ref{Alg:Hybrid_Precoding2}, respectively, when the total transmit power is $25$\,dBm. We can see that both algorithms converge to a stationary point after five iterations. 
The objective function value after convergence is smaller in Algorithm~\ref{Alg:Hybrid_Precoding} than in Algorithm~\ref{Alg:Hybrid_Precoding2} due to the additional constraint on the digital precoder entries in Algorithm~\ref{Alg:Hybrid_Precoding2}, imposed by fronthaul limitations. This constraint limits the digital precoder entries to a few possible values, which also explains why Algorithm~\ref{Alg:Hybrid_Precoding} achieves a higher average sum rate after convergence.

\begin{figure}[t]
    \centering
    \includegraphics[width =0.86 \columnwidth]{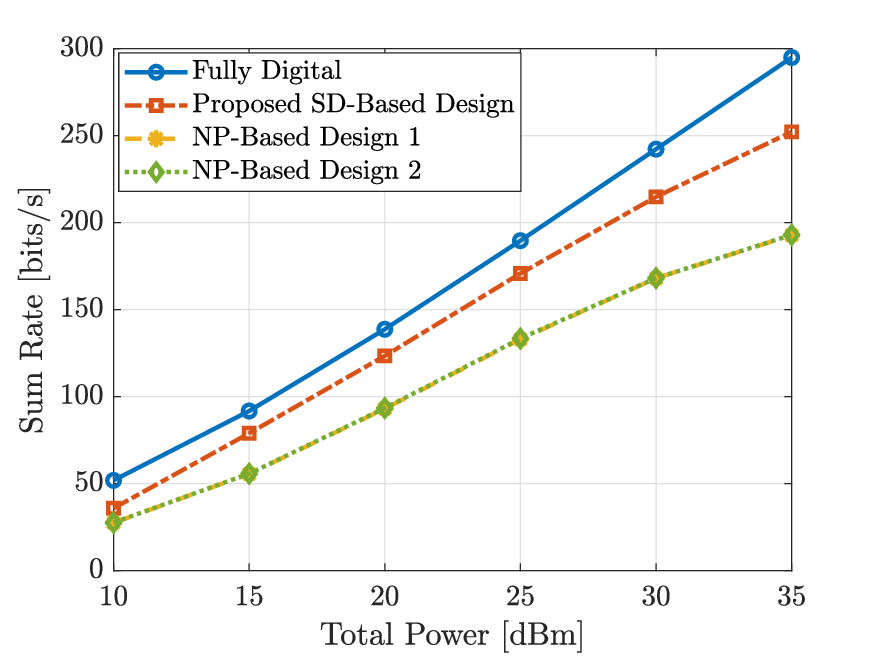}
    \caption{Sum rate of the fully digital precoder and three hybrid precoders with different analog precoder designs.} 
    \label{fig:rate1}
    \vspace{-5mm}
\end{figure}

We now evaluate the sum rate performance by comparing Algorithm~\ref{Alg:Hybrid_Precoding} with two benchmark methods that use nearest point mapping, a common approach for low-resolution analog precoder design \cite{Sohrabi2016Hybrid,ni2017near,Chen2018Low}. These benchmarks assume infinite-resolution phase shifters, then map each analog precoder entry to the nearest value in $\mathcal{D}$ after convergence.
In the first benchmark, named ``NP-Based Design 1'' in Fig.~\ref{fig:rate1}, the infinite-resolution solution to problem \eqref{eq:analog_prec_optimization} is obtained as  $\vect{F}^{\mtr{inf}}_{\mtr{RF}} = \mtr{exp} \left(\imagunit\arg(\vect{F}_{\mtr{FD}}^\star \vect{F}_{\mtr{BB}}^\dagger)\right)$,
where $\vect{F}_{\mtr{BB}}^\dagger$ denotes the pseudoinverse of $\vect{F}_{\mtr{BB}}$. The second benchmark, labeled as ``NP-Based Design 2'' in Fig.~\ref{fig:rate1}, follows the method proposed in \cite{ni2017near} for solving problem \eqref{eq:analog_prec_optimization}.
Fig.~\ref{fig:rate1} shows the sum rate versus the transmit power with an analog phase shifter resolution of $b = 1$. The proposed SD-based design outperforms the two benchmarks based on nearest point mapping by optimizing the low-resolution analog precoder in each iteration. The fully digital design without fronthaul limitation performs best due to its greater control over amplitude and phase and better multi-user interference suppression.

\begin{figure}[t]
    \centering
    \includegraphics[width = 0.86\columnwidth]{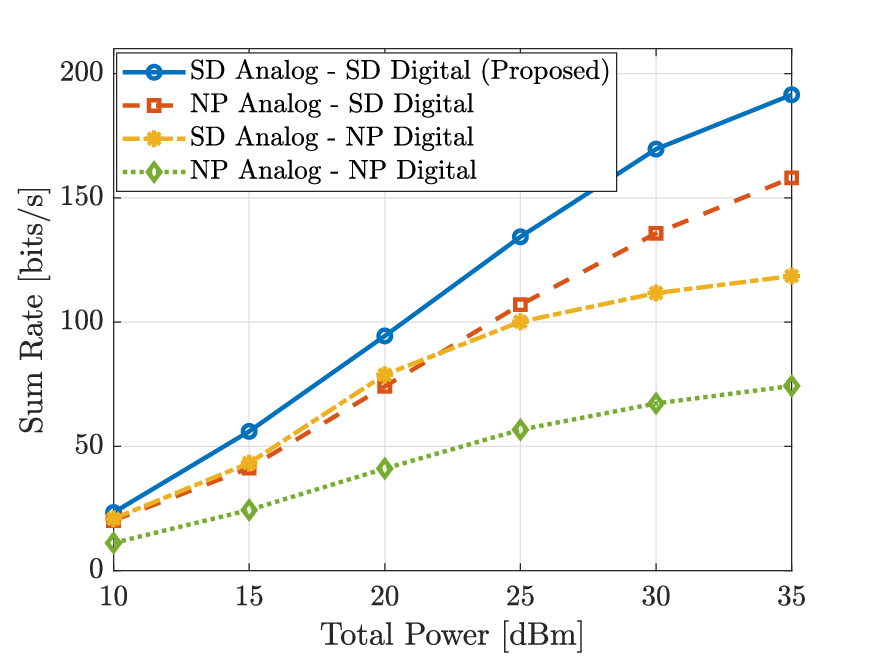}
    \caption{Sum rate of the SD-based analog and digital precoders, and three schemes based on nearest point mapping.   } 
    \label{fig:rate2}
    \vspace{-5.9mm}
\end{figure}

Next, we evaluate the performance of the proposed hybrid precoder optimization in Algorithm~\ref{Alg:Hybrid_Precoding2}, considering fronthaul capacity limitations. Fig.~\ref{fig:rate2} shows the sum rate versus the transmit power for four analog and digital precoder designs, with bit resolutions set as  
$b=q=1$. The solid blue curve represents the proposed SD-based design in Algorithm~\ref{Alg:Hybrid_Precoding2}, while other curves show designs using nearest point mapping for either one or both of the precoders. The SD-based design outperforms the benchmarks, with a growing performance gap as the transmit power increases, because it better approximates the fully digital precoder and thereby can handle interference more effectively. The scheme with only SD-based analog precoder performs slightly better than the one with only SD-based digital precoder at the beginning; however, at higher transmit powers, the SD-based digital precoder significantly outperforms the SD-based analog precoder. This highlights the importance of optimal digital precoder design at high SNRs, as the digital precoder controls both the phase and amplitude of the transmitted signal, while the analog precoder controls only the phase. 

\section{Conclusions}
In this paper, we addressed the key precoding challenges with low-resolution phase shifters and precoding quantization over a limited capacity fronthaul in hybrid multi-carrier MIMO systems. Using a matrix decomposition approach,  we minimized the Euclidean distance between an optimized fully digital high-resolution precoder and the finite-resolution hybrid precoder.
More specifically, we utilized the SD algorithm to design the precoders to minimize the performance degradation caused by the low resolution of the phase shifters and the fronthaul link. Our numerical simulations confirmed that the proposed designs outperform the widely adopted benchmarks based on nearest point mapping.

\bibliographystyle{IEEEtran}
\bibliography{refs} 
\end{document}